\documentclass[final, 11pt, times, twocolumn]{elsarticle}

\usepackage{amssymb}
\usepackage{xcolor}
\usepackage[normalem]{ulem}

\usepackage{geometry}
\newcommand{\reply}[1]{#1}

\journal{Science Bulletin {\rm as a Perspective article}}

\begin{document}

\begin{frontmatter}



\title{Neutron stars as extreme laboratories  for gravity tests}


\author[inst1,inst2]{Lijing Shao}
\ead{lshao@pku.edu.cn}
\affiliation[inst1]{organization={Kavli Institute for Astronomy and Astrophysics},
            addressline={Peking University}, 
            city={Beijing},
            postcode={100871}, 
            country={China}}

\affiliation[inst2]{organization={National Astronomical Observatories},
            addressline={Chinese Academy of Sciences}, 
            city={Beijing},
            postcode={100012}, 
            country={China}}

\author[inst3]{Kent Yagi\fnref{editornote}}
\fntext[editornote]{These authors contributed equally to this work.}
\affiliation[inst3]{organization={Department of Physics},
            addressline={University of Virginia}, 
            city={Charlottesville},
            postcode={22904-4714}, 
            state={Virginia},
            country={United States}}

\begin{abstract}
\reply{Neutron stars are versatile in their application to studying various
important aspects of fundamental physics, in particular  strong-field gravity
tests and the equation of state for super-dense nuclear matter at low
temperatures.  }However, in many cases these two objectives are degenerate to
each other. We discuss how pulsar timing and gravitational waves provide
accurate measurements of neutron star systems and how to effectively break the
degeneracy using tools like universal relations. \reply{We also present
perspectives on future opportunities and challenges in the field of neutron star
physics.}   
\end{abstract}


\end{frontmatter}




Neutrons were discovered 90 years ago by James Chadwick. The concept of neutron
stars was hypothesized around that time by Lev Landau, Walter Baade, and Fritz
Zwicky, and \reply{it was} further developed  by Richard Tolman, Robert
Oppenheimer, George Volkoff, and other physicists. Neutron stars are
astrophysical compact objects formed after the death of massive stars (more
massive than our own Sun).  \reply{According to current understanding, the
typical mass of neutron stars is comparable to {that of} the Sun, while the
radius is only $\sim 10$\,km. As its name {implies}, the main interior
ingredients of such stars are neutrons, but they also consist of protons,
electrons, muons, and  {presumably} even more exotic particles, like hyperons,
kaons, or quarks.} The field of neutron stars was \reply{bolstered} by the
discovery of radio pulsars by Jocelyn Bell Burnell, Antony Hewish, and
collaborators in 1967~\cite{Hewish:1968bj}. A famous achievement \reply{brought
forth} by the first binary pulsar, the so-called Hulse-Taylor pulsar, is the
first-ever empirical proof that gravitational waves exist in our
Universe~\cite{Taylor:1979zz}. In 2017, the first binary neutron star merger was
observed directly via gravitational waves (ripples of spacetime) and accompanied
by enormous electromagnetic follow-up
observations~\cite{LIGOScientific:2017vwq}. This event, known as GW170817,
marked the dawn of multi-messenger astronomy and was chosen as ``Science's 2017
Breakthrough of the Year.'' 

Studies of neutron stars provide us with unique extreme laboratories for
fundamental physical laws, including tests of gravitational theories in the
strong field, \reply{superdense nuclear {matter} at low {temperatures}}, and
energetic astrophysical phenomena underpinning the evolution of stars. Interests
in the research field of neutron stars range from gravity,  nuclear and particle
physics, plasma physics, and even condensed matter physics. \reply{From an
observational point of view, pulsar-timing and gravitational-wave observations
provide  two widely used tools to achieve these rich {scientific} goals.} 


Pulsar timing provides the most accurate measurements in modern astronomy.
\reply{It is usually done by large-area radio telescopes or  {arrays and}
sometimes {done} by high-energy satellites.} These instruments record the times
of arrival of pulses originating from rotating neutron stars {that emit} beams
towards Earth.  These times of arrival are calibrated with atomic clocks at
telescope {sites,} and they form the central data  {for} a clean pulsar-timing
experiment. Such an experiment can last for decades.  For a neutron star in a
binary system that is observed as a pulsar on Earth, the times of arrival, after
subtracting the contribution from {the} interstellar medium and {the}
telescopes' motion, foremost reflect the movement of the pulsar. \reply{As per
the gravitational dynamics of the binary, information about underlying gravity
is encoded with the pulsar's movements and hence with the times of arrival.} By
{reverse}   engineering, we infer the underlying gravity theory and check if it
is consistent with predictions from General Relativity, the {\it tour-de-force}
of Albert Einstein's masterpiece describing gravitation and spacetime. 

Why do we need to test General Relativity? \reply{From a theoretical
{standpoint}, gravity cannot be correctly quantized under General Relativity to
describe microcosmic-scale gravitational physics and {to} formulate a theory to
unify all four known forces in physics.} {From an observational standpoint}, we
only know about 5\% of the energy content of matter in our Universe, and the
remaining 95\% consists of dark matter and dark energy. The former is
responsible for the missing mass observed within galaxies and beyond, while the
latter, having negative pressure, is the origin of \reply{the currently
accelerating expansion of the Universe}.  Interestingly, one could explain these
cosmological unsolved problems by going beyond General Relativity without
introducing unknown dark matter or dark energy.

Binary pulsars are one of the best testbeds for gravitational
experiments~\cite{Damour:2007uf, Shao:2016ezh}. {For} the Hulse-Taylor pulsar
PSR~B1913+16, astronomers have measured the shrinkage of the orbit  {due to} the
emission of gravitational waves which takes away reservoir energy  from the
orbit~\cite{Taylor:1979zz}.  \reply{This was the first indirect sign that
gravitational waves exist  {and,} at that {time,} greatly reinforced the
scientific motivation to build gravitational wave detectors, which  {have proven
to be} extremely successful at the present time.}    However, nowadays the
Hulse-Taylor pulsar is no longer competitive in measuring the orbital shrinkage
as the uncertain contamination from its position and motion in the Galactic
potential has dominated the error budget.  Fortunately, we have several dozens
of pulsar systems to continue performing tests of gravity, with the double
pulsar PSR~J0737$-$3039 being a \reply{prime} example~\cite{Kramer:2021jcw}.

\begin{figure}[!h]
    \centering
    \includegraphics[width=\linewidth]{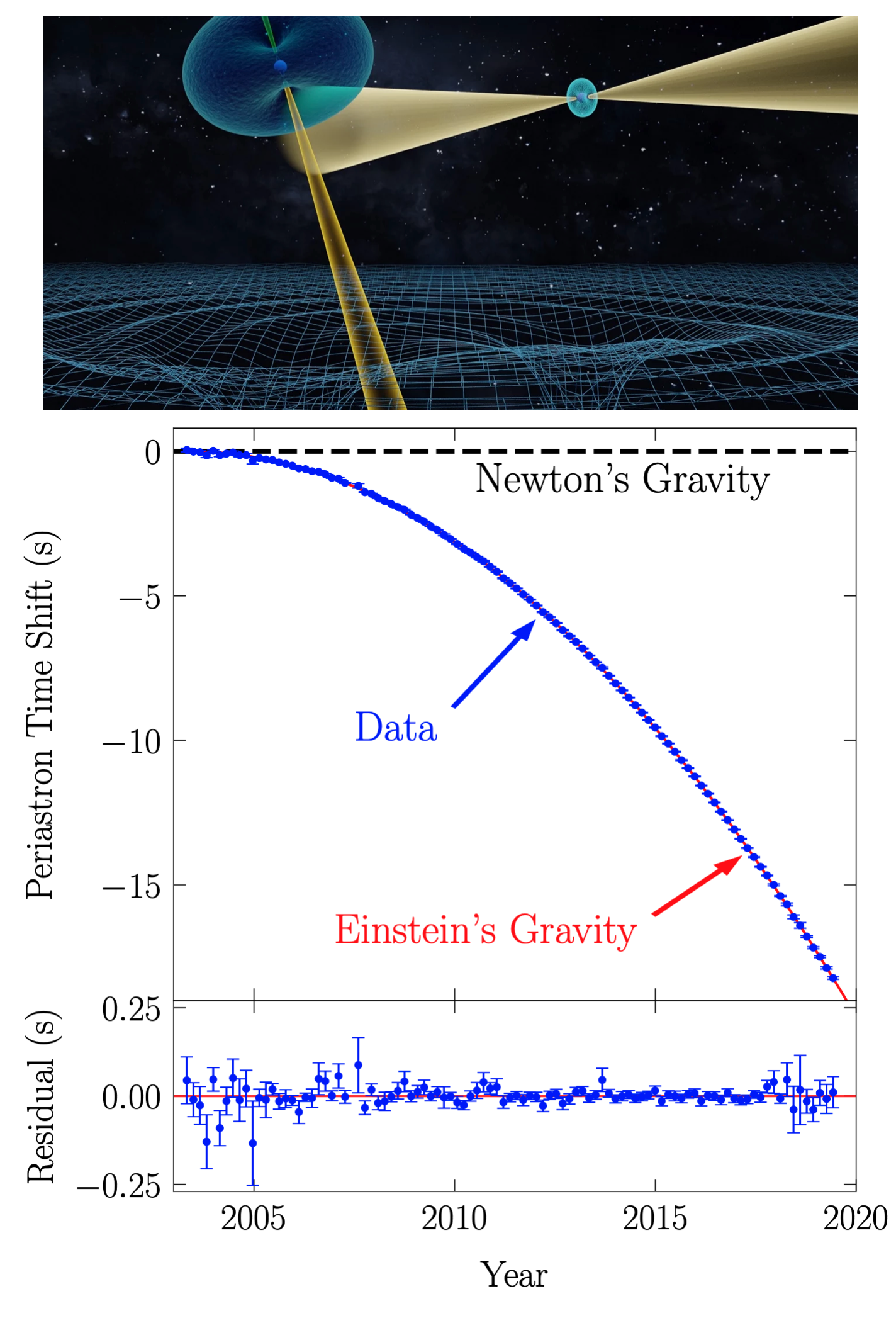}
    \caption{\label{fig:dp_para} ({\it Upper}) Artistic illustration of the
    double pulsar (figure courtesy of M.\ Kramer). ({\it Lower}) The top panel
    presents the confrontation of the double pulsar data with the binary
    periastron time shift due to gravitational-wave radiation predicted in
    Newton's gravity and Einstein's gravity~\cite{Kramer:2021jcw}. The  bottom
    panel shows the deviation from Einstein's gravity, which is consistent with
    random noises, with a normalized $\chi^2 \simeq 0.76$.
    }
\end{figure} 

The double pulsar is a unique system that has two neutron stars, both of which
were observed as active pulsars (see Fig.~\ref{fig:dp_para}). It has a
relativistic orbit with an orbital period $P_{\rm b} = 0.102 251 559 297 3 \pm
0.000 000 000 0010$ day. Combined with its exquisite measurement accuracy, a
number of relativistic gravity tests become available within this system,
including higher-order effects in the orbital motion, light propagation in a
strong gravitational field, and so on.  The double pulsar also revealed the
first-ever information \reply{about the moment of inertia of  {a single} neutron
star} via its effects on the orbit. The moment of inertia is a quantity
characterizing how easily one can spin an object (figure skaters change their
moment of inertia by stretching and shrinking their arms to change the rate of
their spin on ice)  and is related to certain universal relations that we will
explain later. The most valuable fundamental test comes from the measurement of
{the} double pulsar's orbital decay, whose precision exceeds all existing
results by orders of magnitude. As shown in Fig.~\ref{fig:dp_para}, while
Newton's gravity theory obviously fails the test, Einstein's General Relativity
gives a precise prediction of the data and passes the test with flying
colors~\cite{Kramer:2021jcw}. It is worth mentioning {that} such a test of
gravitational radiation in General Relativity has a much broader range of
implications in fundamental theories. A few modified gravity theories are
constructed to account for  dark-matter and/or dark-energy phenomena by
introducing extra degrees of freedom, which are very likely to result in new
channels of gravitational radiation. Thus, the radiative test from the orbital
decay of the double pulsar has strong implications for the above cosmological
problems.

Gravity tests from radio pulsars are versatile~\cite{Shao:2016ezh}. The internal
strong gravity of neutron stars additionally promotes these tests to the strong
gravity field, meaning that neutron stars also probe gravitational phenomena
which are unique to spacetime regions where  gravity is  {extreme} . For
example, one of the most well-studied alternative theories of gravity is
scalar-tensor theory, where gravity is coupled to a scalar field. Just like
{how} electric charges source electric fields, there are sources called scalar
charges that generate scalar fields in these theories beyond General Relativity.
A phase-transition-like phenomenon called ``scalarization'' may arise where a
large number of scalar charges are spontaneously produced for suitable neutron
stars within a certain mass range~\cite{Damour:2007uf}.  Neutron stars appear to
be a natural laboratory for testing these strong-field
predictions~\cite{Shao:2017gwu}

Pulsar timing is quite complementary to the novel strong-field tests of gravity
with gravitational waves and black hole shadows that have become available in
the last several years. First, compared with short gravitational-wave bursts
from stellar compact binary coalescences, pulsar timing is a long-term
experiment. Therefore, while the former probes highly dynamical spacetimes and
has a greater sensitivity to higher-order effects beyond Newtonian gravity, the
latter probes quasi-stationary spacetimes and has \reply{a higher sensitivity to
effects that accumulate  {over} time.} Both aspects are equally important to
scrutinize a gravitational theory. Second, gravitational-wave observations and
black hole shadows are superb means to study black hole spacetimes, while
pulsars are especially powerful to investigate a few gravitational theories
\reply{where the  {coupling} between matter and {spacetime is} important.} The
aforementioned neutron star scalarization is a vivid example along this
line.\footnote{Gravitational waves from binary neutron star mergers are also
useful to probe  {the} dynamical aspect of neutron star scalarization.}
\reply{Third, pulsar timing and black hole shadows not only deal with the
dynamics under gravity,  but also study the propagation of {light} in curved
spacetimes.} In contrast, gravitational waves handle nonlinear back reactions
of,  \reply{in quantum parlance, {gravitons} in spacetime.}  \reply{Overall,
concerning the complementarities of pulsar timing, gravitational waves, and
black hole shadows, combinations of these three kinds of strong-gravity
experiments were shown to be {powerful} and would allow us to understand
gravitation and spacetime more {thoroughly}~\cite{Shao:2017gwu,
EventHorizonTelescope:2022xqj}.} 

\begin{figure}[h]
    \centering
\includegraphics[width=\linewidth, trim={3.7cm 0 3.7cm 0}, clip]{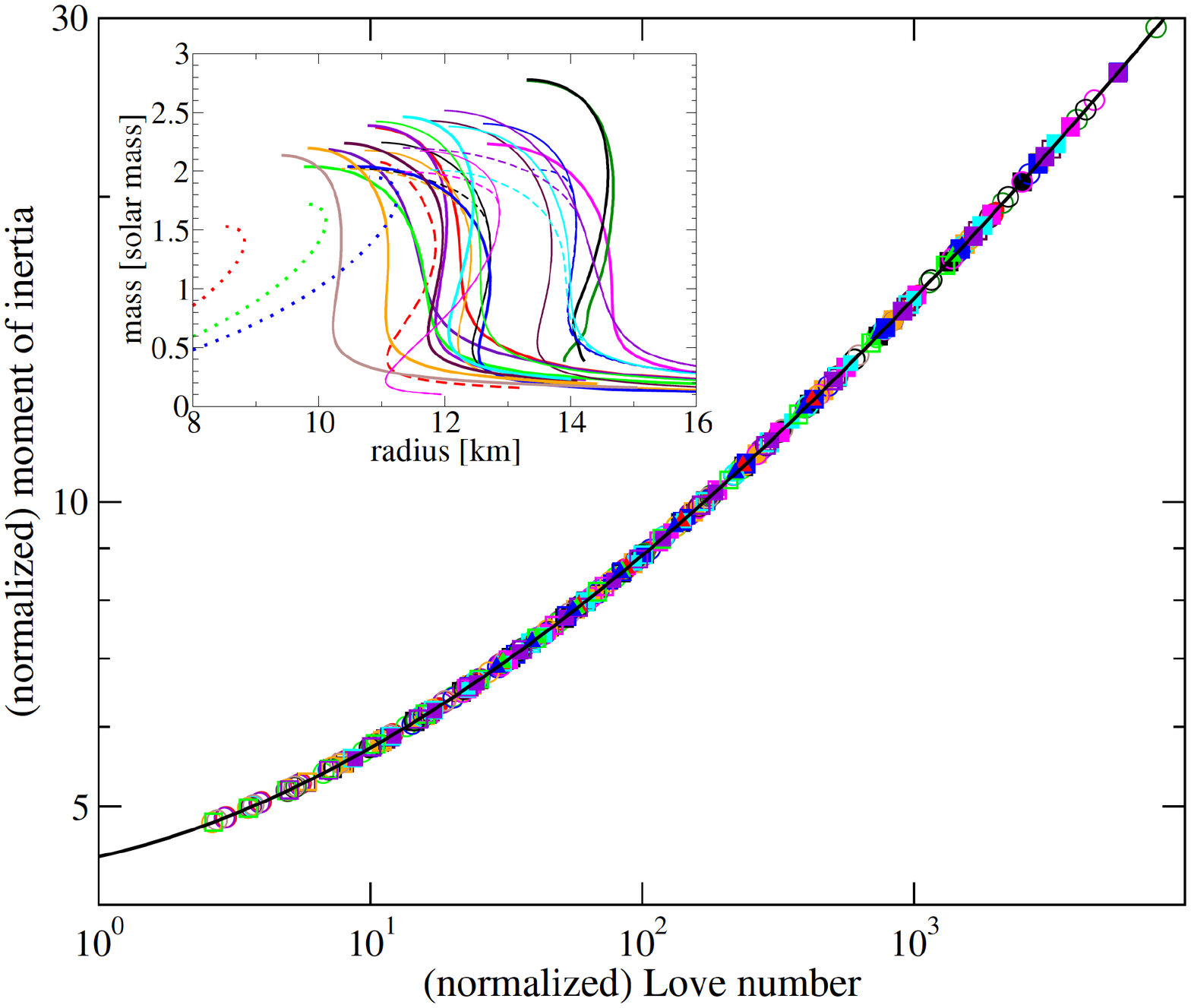}
\caption{\label{fig:I-Love-MR} ({\it Main}) Universal relation between the
moment of inertia and Love number for neutron stars (I-Love relation). Each
quantity has been normalized to be dimensionless using the mass, speed of light
$c$, and gravitational constant $G$. Each marker corresponds to one equation of
state and the solid curve is the fit. ({\it Inset}) The relation between the
mass and radius that strongly depends on the equations of state (shown by
different colors). Figure taken and modified from Ref.~\cite{Yagi:2016bkt}.
}
\end{figure}

Due to their strong gravitational field and extreme density, neutron stars are
ideal astrophysical laboratories to probe not only gravitational physics but
also nuclear physics. For example, the relation between the neutron star mass
and radius depends very sensitively on the underlying equations of state of
nuclear matter, which are relations between pressure and density (see the inset
in Fig.~\ref{fig:I-Love-MR}). Heavy pulsars with masses $\sim 2\, M_\odot$ have
ruled out some of the ``soft'' equations of state whose maximum mass of a
neutron star is smaller than $2\, M_\odot$. Recent x-ray observations with the
NICER satellite have measured the mass and radius of neutron stars
independently, which \reply{allowed} us to constrain the equations of state. The
LIGO/Virgo Collaboration has measured the tidal deformability of neutron stars
in a merging binary, which has \reply{further} constrained the equations of
state. The aforementioned measurements are not precise enough at present, and we
still have sizable uncertainties in the equations of state.

This means we face a challenge that if one wants to test strong-field gravity,
there will be systematic errors due to uncertainties in nuclear physics.  For
example, the mass-radius relation of neutron stars depends not only on the
nuclear equations of state but also on gravitational physics.  Therefore, if we
can measure the mass and radius of a neutron star simultaneously, we can probe
gravitational {theory,} provided that we know the underlying equations of state.
If, on the other hand, we do not have a complete understanding of the equations
of state, there will be systematic uncertainties in the probes of gravitational
physics.

One way to break such a degeneracy between uncertainties in gravitational and
nuclear physics is to use universal relations among neutron star observables
that do not depend sensitively on the equations of state. One example of such
relations is the ``I-Love-Q'' relation between the moment of inertia (typically
denoted as $I$), tidal Love number (or tidal {deformability, which} tells us how
easily an object can tidally deform), and quadrupole moment (characterizing a
quadrupolar deformation of an object in its shape, typically represented by
$Q$)~\cite{Yagi:2013bca} (see Fig.~\ref{fig:I-Love-MR}). Although these
relations are almost independent of the choice of the equations of state
\reply{(with an equation-of-state variation of $\sim 1\%$ or even smaller)},
they do depend on the underlying gravitational physics. Therefore, if we can
independently measure any two of the I-Love-Q trio, we can \reply{test gravity
theories without  {the contamination from} uncertainties in nuclear physics.} In
this regard, the I-Love relation may be the most interesting because the Love
number has been measured by gravitational waves from colliding neutron
stars~\cite{LIGOScientific:2017vwq} while radio observations of the double
pulsar binary have started to constrain the moment of inertia of the primary
pulsar as mentioned earlier~\cite{Kramer:2021jcw}.  Yagi and
Yunes~\cite{Yagi:2013bca} applied this idea to test a specific theory of gravity
that breaks parity (mirror symmetry) and showed that multi-messenger
observations of neutron stars can be probed with an accuracy that is several
orders of magnitude higher than the current bound from Solar System experiments.
This was recently confirmed by Silva \textit{et al.}~\cite{Silva:2020acr} who
used the direct measurement of {the} Love number with gravitational waves from
LIGO/Virgo detectors and inferred measurement of {the} moment of inertia from
x-ray observations with the NICER satellite. \reply{The equation-of-state
variation in the I-Love-Q relations is small enough to perform the above tests
of gravity given that typical measurement errors of I-Love-Q are of
$\mathcal{O}(10\%)$.}

This avenue of using universal relations to test strong-field gravity seems
promising, though there are still \reply{some issues (or questions) that need to
be investigated further}. First, universal relations are typically studied for a
single neutron star. This means that they can be applied to observations of two
neutron stars with similar masses, but may fail for neutron stars with different
masses. Therefore, one needs to study whether universal relations hold for
observables of two neutron stars with different masses. \reply{A first attempt
was made by Saffer and Yagi~\cite{Saffer:2021gak}, but a more thorough analysis
is necessary. }

Second, there are many other universal relations known within General
Relativity, so one needs to study their applications on strong-field tests of
gravity. One example is the relation between the Love number and an oscillation
frequency of a neutron star (called the fundamental mode oscillation, also known
as the f-mode oscillation)~\cite{Chan:2014kua}. The Love number has already been
measured through static tidal effects, while the f-mode oscillation is expected
to be detected with future gravitational-wave observations through dynamical
tides. The advantage of using this relation is that \reply{both of the
quantities can be observed with gravitational waves} from a colliding binary
neutron star. Therefore, we do not suffer from the issue mentioned in the
previous paragraph  {of} using universal relations for neutron stars with
different masses. \reply{On the other hand, when the masses of neutron stars are
similar, say $\sim 1.4 \, M_\odot$, then we can Taylor-expand each observable
(that is a function of the mass) relevant for the universal relations (like the
Love number or f-mode oscillation frequency) about the fiducial mass, like $1.4
\, M_\odot$, and we can measure the leading quantity in the expansion that
corresponds to each quantity evaluated at $1.4 \, M_\odot$. Since this quantity
is common to all neutron stars with masses similar to the fiducial mass, we can
combine different observations to improve the measurement accuracy of such
observable (by a factor of $\sqrt{N}$ where $N$ is the number of events).}

Third, most previous studies on tests of gravity with universal relations
focused on probing specific gravitational theories. However, given that there
are numerous modified theories of gravity being proposed, testing each theory
one after another would be extremely time-consuming.  A more efficient approach
would be to first perform a model-independent test without specifying a theory
and then map the information to a specific theory. One way of performing such a
model-independent test is to construct parameterized universal relations that
are characterized by parameters representing non-General-Relativity effects. If
a known mapping exists between such parameters and theoretical constants in each
theory beyond General Relativity, one can easily probe/constrain the theory.
This approach was proposed and studied by Silva \textit{et
al.}~\cite{Silva:2020acr} for the I-Love relation, taking the parity-breaking
theory mentioned earlier as a base theory for constructing the parameterized
relation. This is a very important  {attempt,} and a more systematic analysis
needs to be carried out for other universal relations and with other theories to
find the mapping.

\reply{Since the inception of the concept of neutron stars about 90 years ago,
they have been playing important roles in various aspects of fundamental
physics, particularly in testing gravitational theories and probing dense
nuclear  matter.} Realizing that neutron stars are very tiny celestial objects
(with a size smaller than a typical city) and they are very distant from Earth
(with pulsars being thousands of light-years away and binary neutron star
mergers being more than millions of light-years away), it is truly remarkable
that tons of precious information have been obtained from them. The future of
this field is extremely bright. For pulsar observations, the next-generation
radio telescope, the Square Kilometre Array~\cite{Bull:2018lat}, will probe the
Southern sky with unprecedented sensitivity. A census of radio pulsars in the
Milky Way is to be made, probably resulting in discoveries like {pulsar-black
hole} binaries. For gravitational waves, the third generation of detectors are
under construction~\cite{Kalogera:2021bya}, and they will detect almost all
binary neutron star mergers in the Universe. With useful tools like the
universal relations at hand, astrophysicists will  disentangle uncertainties in
tests of gravity and nuclear  {matter} and eventually advance knowledge greatly
in both subjects.

\section*{Acknowledgements}

\reply{We thank Sid Ajith for carefully proofreading the manuscript.} L.S.\ was
supported by the National SKA Program of China (2020SKA0120300),  the National
Natural Science Foundation of China (11975027, 11991053, 11721303), and the Max
Planck Partner Group Program funded by the Max Planck Society.  K.Y.\
acknowledges support from NSF Grant PHY-1806776, a Sloan Foundation Research
Fellowship and the Owens Family Foundation.  K.Y. would like to also acknowledge
support by the COST Action GWverse CA16104 and JSPS KAKENHI Grants No.
JP17H06358.

 \bibliographystyle{apsrev} 
 \bibliography{refs}

\begin{thebibliography}{15}
\expandafter\ifx\csname natexlab\endcsname\relax\def\natexlab#1{#1}\fi
\expandafter\ifx\csname bibnamefont\endcsname\relax
  \def\bibnamefont#1{#1}\fi
\expandafter\ifx\csname bibfnamefont\endcsname\relax
  \def\bibfnamefont#1{#1}\fi
\expandafter\ifx\csname citenamefont\endcsname\relax
  \def\citenamefont#1{#1}\fi
\expandafter\ifx\csname url\endcsname\relax
  \def\url#1{\texttt{#1}}\fi
\expandafter\ifx\csname urlprefix\endcsname\relax\def\urlprefix{URL }\fi
\providecommand{\bibinfo}[2]{#2}
\providecommand{\eprint}[2][]{\url{#2}}

\bibitem[{\citenamefont{Hewish et~al.}(1968)\citenamefont{Hewish, Bell,
  Pilkington, Scott, and Collins}}]{Hewish:1968bj}
\bibinfo{author}{\bibfnamefont{A.}~\bibnamefont{Hewish}},
  \bibinfo{author}{\bibfnamefont{S.~J.} \bibnamefont{Bell}},
  \bibinfo{author}{\bibfnamefont{J.~D.~H.} \bibnamefont{Pilkington}},
  \bibinfo{author}{\bibfnamefont{P.~F.} \bibnamefont{Scott}}, \bibnamefont{and}
  \bibinfo{author}{\bibfnamefont{R.~A.} \bibnamefont{Collins}},
  \bibinfo{journal}{Nature} \textbf{\bibinfo{volume}{217}},
  \bibinfo{pages}{709} (\bibinfo{year}{1968}).

\bibitem[{\citenamefont{Taylor et~al.}(1979)\citenamefont{Taylor, Fowler, and
  McCulloch}}]{Taylor:1979zz}
\bibinfo{author}{\bibfnamefont{J.~H.} \bibnamefont{Taylor}},
  \bibinfo{author}{\bibfnamefont{L.~A.} \bibnamefont{Fowler}},
  \bibnamefont{and} \bibinfo{author}{\bibfnamefont{P.~M.}
  \bibnamefont{McCulloch}}, \bibinfo{journal}{Nature}
  \textbf{\bibinfo{volume}{277}}, \bibinfo{pages}{437} (\bibinfo{year}{1979}).

\bibitem[{\citenamefont{Abbott et~al.}(2017)}]{LIGOScientific:2017vwq}
\bibinfo{author}{\bibfnamefont{B.~P.} \bibnamefont{Abbott}}
  \bibnamefont{et~al.} (\bibinfo{collaboration}{LIGO Scientific, Virgo}),
  \bibinfo{journal}{Phys. Rev. Lett.} \textbf{\bibinfo{volume}{119}},
  \bibinfo{pages}{161101} (\bibinfo{year}{2017}), \eprint{1710.05832}.

\bibitem[{\citenamefont{{Damour}}(2009)}]{Damour:2007uf}
\bibinfo{author}{\bibfnamefont{T.}~\bibnamefont{{Damour}}}, in
  \emph{\bibinfo{booktitle}{{Physics of Relativistic Objects in Compact
  Binaries: From Birth to Coalescence}}}, edited by
  \bibinfo{editor}{\bibfnamefont{M.}~\bibnamefont{Colpi}},
  \bibinfo{editor}{\bibfnamefont{P.}~\bibnamefont{Casella}},
  \bibinfo{editor}{\bibfnamefont{V.}~\bibnamefont{Gorini}},
  \bibinfo{editor}{\bibfnamefont{U.}~\bibnamefont{Moschella}},
  \bibnamefont{and} \bibinfo{editor}{\bibfnamefont{A.}~\bibnamefont{Possenti}}
  (\bibinfo{publisher}{Springer, Dordrecht}, \bibinfo{year}{2009}), vol.
  \bibinfo{volume}{359}, p.~\bibinfo{pages}{1}, \eprint{0704.0749}.

\bibitem[{\citenamefont{Shao and Wex}(2016)}]{Shao:2016ezh}
\bibinfo{author}{\bibfnamefont{L.}~\bibnamefont{Shao}} \bibnamefont{and}
  \bibinfo{author}{\bibfnamefont{N.}~\bibnamefont{Wex}}, \bibinfo{journal}{Sci.
  China Phys. Mech. Astron.} \textbf{\bibinfo{volume}{59}},
  \bibinfo{pages}{699501} (\bibinfo{year}{2016}), \eprint{1604.03662}.

\bibitem[{\citenamefont{Kramer et~al.}(2021)}]{Kramer:2021jcw}
\bibinfo{author}{\bibfnamefont{M.}~\bibnamefont{Kramer}} \bibnamefont{et~al.},
  \bibinfo{journal}{Phys. Rev. X} \textbf{\bibinfo{volume}{11}},
  \bibinfo{pages}{041050} (\bibinfo{year}{2021}), \eprint{2112.06795}.

\bibitem[{\citenamefont{Shao et~al.}(2017)\citenamefont{Shao, Sennett,
  Buonanno, Kramer, and Wex}}]{Shao:2017gwu}
\bibinfo{author}{\bibfnamefont{L.}~\bibnamefont{Shao}},
  \bibinfo{author}{\bibfnamefont{N.}~\bibnamefont{Sennett}},
  \bibinfo{author}{\bibfnamefont{A.}~\bibnamefont{Buonanno}},
  \bibinfo{author}{\bibfnamefont{M.}~\bibnamefont{Kramer}}, \bibnamefont{and}
  \bibinfo{author}{\bibfnamefont{N.}~\bibnamefont{Wex}},
  \bibinfo{journal}{Phys. Rev. X} \textbf{\bibinfo{volume}{7}},
  \bibinfo{pages}{041025} (\bibinfo{year}{2017}), \eprint{1704.07561}.

\bibitem[{\citenamefont{Akiyama et~al.}(2022)}]{EventHorizonTelescope:2022xqj}
\bibinfo{author}{\bibfnamefont{K.}~\bibnamefont{Akiyama}} \bibnamefont{et~al.}
  (\bibinfo{collaboration}{Event Horizon Telescope}),
  \bibinfo{journal}{Astrophys. J. Lett.} \textbf{\bibinfo{volume}{930}},
  \bibinfo{pages}{L17} (\bibinfo{year}{2022}).

\bibitem[{\citenamefont{Yagi and Yunes}(2017)}]{Yagi:2016bkt}
\bibinfo{author}{\bibfnamefont{K.}~\bibnamefont{Yagi}} \bibnamefont{and}
  \bibinfo{author}{\bibfnamefont{N.}~\bibnamefont{Yunes}},
  \bibinfo{journal}{Phys. Rept.} \textbf{\bibinfo{volume}{681}},
  \bibinfo{pages}{1} (\bibinfo{year}{2017}), \eprint{1608.02582}.

\bibitem[{\citenamefont{Yagi and Yunes}(2013)}]{Yagi:2013bca}
\bibinfo{author}{\bibfnamefont{K.}~\bibnamefont{Yagi}} \bibnamefont{and}
  \bibinfo{author}{\bibfnamefont{N.}~\bibnamefont{Yunes}},
  \bibinfo{journal}{Science} \textbf{\bibinfo{volume}{341}},
  \bibinfo{pages}{365} (\bibinfo{year}{2013}), \eprint{1302.4499}.

\bibitem[{\citenamefont{Silva et~al.}(2021)\citenamefont{Silva, Holgado,
  C\'ardenas-Avenda\~no, and Yunes}}]{Silva:2020acr}
\bibinfo{author}{\bibfnamefont{H.~O.} \bibnamefont{Silva}},
  \bibinfo{author}{\bibfnamefont{A.~M.} \bibnamefont{Holgado}},
  \bibinfo{author}{\bibfnamefont{A.}~\bibnamefont{C\'ardenas-Avenda\~no}},
  \bibnamefont{and} \bibinfo{author}{\bibfnamefont{N.}~\bibnamefont{Yunes}},
  \bibinfo{journal}{Phys. Rev. Lett.} \textbf{\bibinfo{volume}{126}},
  \bibinfo{pages}{181101} (\bibinfo{year}{2021}), \eprint{2004.01253}.

\bibitem[{\citenamefont{Saffer and Yagi}(2021)}]{Saffer:2021gak}
\bibinfo{author}{\bibfnamefont{A.}~\bibnamefont{Saffer}} \bibnamefont{and}
  \bibinfo{author}{\bibfnamefont{K.}~\bibnamefont{Yagi}},
  \bibinfo{journal}{Phys. Rev. D} \textbf{\bibinfo{volume}{104}},
  \bibinfo{pages}{124052} (\bibinfo{year}{2021}), \eprint{2110.02997}.

\bibitem[{\citenamefont{Chan et~al.}(2014)\citenamefont{Chan, Sham, Leung, and
  Lin}}]{Chan:2014kua}
\bibinfo{author}{\bibfnamefont{T.~K.} \bibnamefont{Chan}},
  \bibinfo{author}{\bibfnamefont{Y.~H.} \bibnamefont{Sham}},
  \bibinfo{author}{\bibfnamefont{P.~T.} \bibnamefont{Leung}}, \bibnamefont{and}
  \bibinfo{author}{\bibfnamefont{L.~M.} \bibnamefont{Lin}},
  \bibinfo{journal}{Phys. Rev. D} \textbf{\bibinfo{volume}{90}},
  \bibinfo{pages}{124023} (\bibinfo{year}{2014}), \eprint{1408.3789}.

\bibitem[{\citenamefont{Weltman et~al.}(2020)}]{Bull:2018lat}
\bibinfo{author}{\bibfnamefont{A.}~\bibnamefont{Weltman}} \bibnamefont{et~al.},
  \bibinfo{journal}{Publ. Astron. Soc. Austral.} \textbf{\bibinfo{volume}{37}},
  \bibinfo{pages}{e002} (\bibinfo{year}{2020}), \eprint{1810.02680}.

\bibitem[{\citenamefont{Kalogera et~al.}(2021)}]{Kalogera:2021bya}
\bibinfo{author}{\bibfnamefont{V.}~\bibnamefont{Kalogera}} \bibnamefont{et~al.}
  (\bibinfo{year}{2021}), \eprint{2111.06990}.

\end{thebibliography}





\end{document}